\documentclass[shortnote,twocolumn]{jpsj2}

\def\runauthor{Online-Journal Subcommittee}

\title{%
Statistical Properties of Height of Japanese Schoolchildren
}

\author{Hiroto \textsc{KUNINAKA}
\thanks{E-mail address: kuninaka@phys.chuo-u.ac.jp}, 
Yu \textsc{Mitsuhashi} and Mitsugu \textsc{Matsushita}}

\inst{%
Department of Physics, Chuo University, Kasuga, Bunkyo-ku, Tokyo 112-8551
}

\recdate{\today}

\kword{%
height, lognormal distribution, normal distribution, multiplicative process, 
data analysis
}

\begin{document}
\maketitle
The height and weight are the fundamental indices of human growth, 
which are strongly affected by various factors such as genetic effects, 
race, nutrition, congenital disease, and so on. 
The relationship between the indices and physical conditions 
has been extensively studied to establish a simple measure,  
such as the body mass index (BMI).
\cite{quetelet_rev,khosla,quetelet} 
One of the serious health problems is obesity which causes 
various diseases such as cardiovascular disease and diabetes. 
Thus, investigating the statistical properties of height and weight 
from the viewpoint of the human growth process is important 
for the prevention of adult diseases.


For the statistical property of height, 
it has been commonly believed that the height distribution 
obeys the normal distribution\cite{clemons,ahearn}, 
\begin{equation}\label{normal}
n(x)=\frac{1}{\sqrt{2\pi} \sigma} \exp \left[-\frac{(x-\mu)^2}
{2\sigma^2} \right], 
\end{equation}
where $n(x)$ is the frequency divided by the total number of 
individuals, $x$ is the height, and $\mu$ and $\sigma$ are the average 
and the standard deviation, respectively. If the height distribution 
were approximated by the normal distribution, it would suggest 
that the growth of individuals is governed by the stochastic additive process 
because the sum of independent random variables approximately 
obeys the normal distribution according to the central limit theorem.

On the other hand, the multiplicative process is often introduced to 
explain the growth process of living organisms. 
In the multiplicative process, the growth of living organisms 
is governed by the time evolution, 
\begin{eqnarray}
X_{t+1} &=& \alpha_{t} X_{t}\\
&=& \alpha_{t} \cdot \alpha_{t-1} \cdots \alpha_{0} \cdot X_{0},
\end{eqnarray}
where $X_{t}$ and $\alpha_{t}$ are the size and the growth rate at 
time $t$, respectively. 
If $\alpha_{t}$ is independent of the size $X_{t}$, 
this process is called Gibrat's process, which reproduces 
the lognormal distribution of $X_{t}$ as 
\begin{equation}\label{lognormal}
f(X_{t})=\frac{1}{\sqrt{2\pi\sigma^2} X_{t}} 
\exp \left[ -\frac{[\ln(X_{t}/T)]^2}{2\sigma^2} \right]
\end{equation}
after long-time growth.\cite{lognormal_dist} 
Here, $f(x)$, $\sigma$, and $T$ are the frequency divided by the total 
number of individuals, the standard deviation, and the median, 
respectively. 
In natural and social phenomena, 
the lognormal distribution can be observed in various cases  
such as population distribution of villages in Japan, \cite{sasaki}
fragmentation of glass rods, \cite{ishii}
and so on.
In this short note, we investigate the height distribution of 
Japanese schoolchildren by data analysis. 
Our data analysis is based on the data of the school health survey 
by the ministry of education, 
culture, sports, science and technology, Japan (MEXT).\cite{mext}
This survey is performed yearly for the schoolchildren 
ranging from $5$ to $17$ in age 
who belong to the schools randomly sampled from all the schools in Japan. 
The survey date ranges from April 1 to June 30 in each year. 
In $2006$, the numbers of sampled schools and children 
were $7755$ and $695600$, respectively. 
The data obtained from MEXT are histogram 
data classified into some classes of height. 
In the case of 17-year-old female students in 2006, 
the range of the data, from $130 \rm{cm}$ to $184 \rm{cm}$, 
are splitted into $55$ equal-sized ($1 \rm{cm}$) classes. 
Note that the interpretation of the data are subject to sampling errors 
because of the sample survey. 


The procedure of our analysis is as follows. First, for the data 
of each survey year, we fit the height distribution in each age 
by the lognormal and the normal distributions, respectively,  
by the use of GNUFIT routine implemented within GNUPLOT, 
which is based on the Marquardt-Levenberg algorithm 
for fitting.\cite{nr} 
Second, to determine which distribution can be better approximation, 
we calculate the sum of squared residuals 
from the data classified into $m$ classes as 
\begin{equation}
R_{i}^{2}=\sum_{j=1}^{m} (O^{j}_{i}-E^{j}_{i})^{2},
\end{equation}
where $O^{j}_{i}$ and $E^{j}_{i}$ are the observed and the estimated values 
for the $j$-th height class at each age $i$, respectively. 
Third, from the sums of squared residuals calculated from both 
the lognormal and the normal distributions, we calculate the ratio 
\begin{equation}
\epsilon_{i} =\frac{R^{(LN)2}_{i}}{R^{(N)2}_{i}},
\end{equation}
where $R^{(LN)2}_{i}$ and $R^{(N)2}_{i}$ are the sum of residuals 
calculated from the lognormal and the normal distributions, respectively. 
The height distribution is well approximated 
by the normal distribution when $\log \epsilon_{i}$ is larger than $0$, 
while it is well approximated by the lognormal distribution 
when $\log \epsilon_{i}$ is less than $0$. 
We calculate $\epsilon_{i}$ for each age $i$ to obtain 
the relationship between $\epsilon_{i}$ and age $i$ 
in a given survey year.

\begin{figure}[t]
\begin{center}
\includegraphics[width=0.4\textwidth]{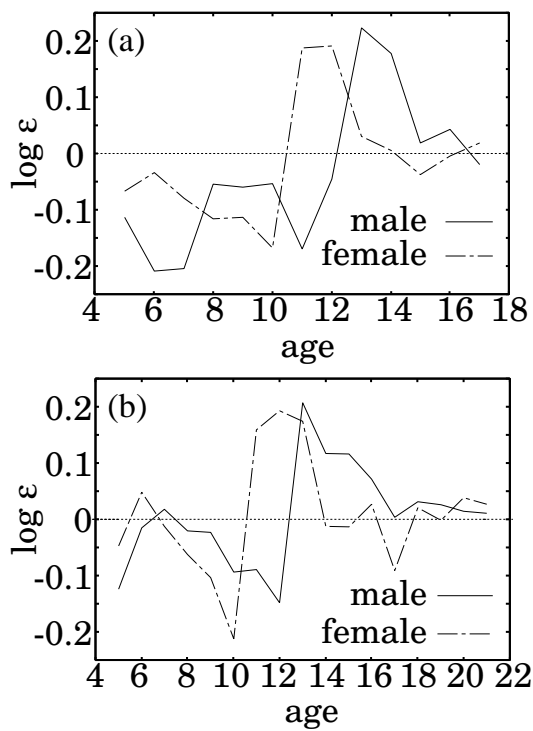}
\end{center}
\caption{
Relationship between age and $\log \epsilon$ in (a) 2006 and (b) 1970. 
Solid and chain lines are for male and female students, respectively.
}
\label{fig2}
\end{figure}

Figure \ref{fig2}(a) shows the relationship between age $i$ and 
$\log \epsilon_{i}$, where $\log$ is common logarithm, 
$\log \equiv \log_{10}$. Here we plot the results obtained from 
the data of both male and female students in 2006. 
We find the clear transition from the lognormal distribution 
to the normal distribution in both male and 
female students around $11$ or $13$ years old. 
The timing of the transition is earlier in female students 
rather than that in male students. 
We commonly call this period puberty in which physical changes occur 
in a child's body to become an adult body capable of reproduction. 
For the adult, the height distribution seems to be approximated by both 
the lognormal and the normal distributions. Limpert {\it et al.} 
pointed out that the similar tendency can be seen in the height distribution 
of female adults.\cite{limpert} 

In the case of another year, Fig. \ref{fig2}(b) shows 
the result in 1970, where 
we also find clear transition from the lognormal distribution to 
the normal distribution although the timing of the transition 
is slightly different from that in 2006. 
These results would imply that the transition from the lognormal distribution 
to the normal distribution in height distribution of schoolchildren occurs 
in any period in Japan. 

Next, we investigate the time evolution of the height distribution 
of the cohort born in 1990. We analyze the data ranging from 1995 to 
2007. Notice that all the members belonging to the cohort are totally 
different because the sampled schools and individuals are different 
according to the survey year. 
\begin{figure}[t]
\begin{center}
\includegraphics[width=0.4\textwidth]{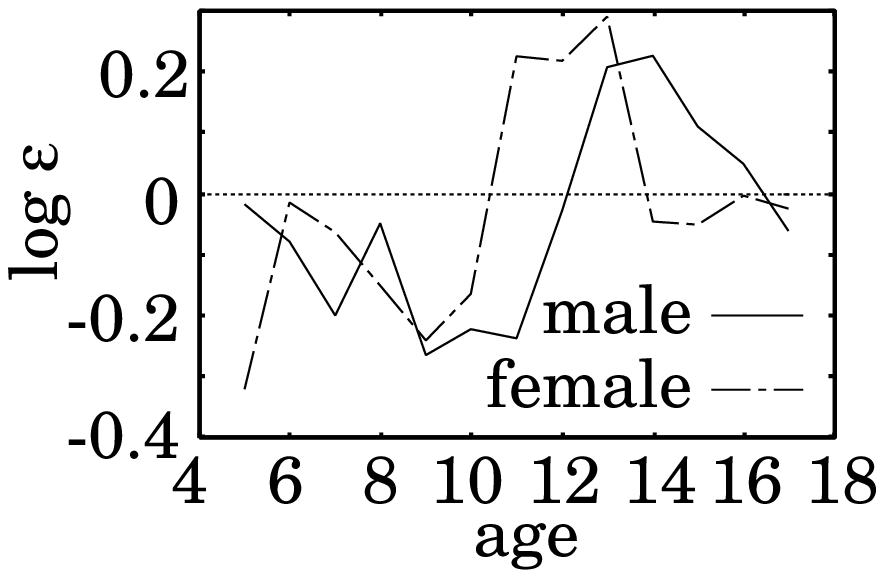}
\end{center}
\caption{
Relationship between age and $\log \epsilon$ for the 
cohort born in 1990. 
Solid and chain lines are for male and female students, respectively.
}
\label{fig3}
\end{figure}
Figure \ref{fig3} shows the relationship between $\log \epsilon$ and 
age of the male and female schoolchildren who belong to the cohort. 
Also in this case, we can observe the similar transition from 
the lognormal distribution to the normal distribution.  
After the transition occurs, 
around 16 years old, the data can fit both distributions. 

From these analyses, we conjecture that the growth process of individuals 
obeys the multiplicative process before puberty. When an individual reaches 
the age around puberty, the growth process becomes rather additive, 
which might cause the transition of the height distribution to the normal 
distribution. Modeling the growth process to reproduce such a transition 
is our near future task. Besides the modeling, we need to analyze the 
data of other countries to investigate whether similar tendency can be 
universally observed among schoolchildren. 
In addition, the growth process of height is closely related to that of 
weight. We will report our analysis of the weight distribution in 
another occasion.

\acknowledgement
HK would like to thank K. Kuninaka and M. Niidome for their 
useful advices. This work is supported by a Grant-in-aid from 
MEXT, Japan (Grant No. 18340115).




\end{document}